\documentclass[11pt]{article}
\usepackage[T1]{fontenc}
\usepackage[utf8]{inputenc}
\usepackage{mathptmx}
\usepackage[letterpaper,margin=1in]{geometry}
\usepackage{amsmath}
\usepackage{graphicx}
\usepackage{booktabs}
\usepackage{array}
\usepackage{tabularx}
\usepackage{enumitem}
\usepackage{float}
\usepackage[font=small,labelfont=bf]{caption}
\usepackage[square,authoryear]{natbib}
\usepackage{xcolor}
\usepackage{microtype}
\usepackage{url}
\usepackage[colorlinks=true,linkcolor=black,citecolor=black,urlcolor=blue!55!black]{hyperref}
\hypersetup{pdftitle={How Long Is the Journey to Work? One-Way Commuting Time in the United States from American Community Survey Data, 2006--2024, with Travel-Diary, European and English Comparisons},pdfauthor={Marcus Oyelaran; Daniel R. Whitcombe},
  pdfsubject={Descriptive analysis of the usual one-way travel time to work of US commuters by year, sex, trip length, mode, departure time, state and metro area, with travel-diary distances and speeds and comparisons with Europe and England. Travel Statistics Bureau preprint.},pdfkeywords={commuting, travel time to work, american community survey, mode choice, working from home, travel survey, descriptive statistics}}
\title{\textbf{How Long Is the Journey to Work? One-Way Commuting Time in the United States from American Community Survey Data, 2006--2024, with Travel-Diary, European and English Comparisons}}
\author{Marcus Oyelaran\textsuperscript{1} and Daniel R. Whitcombe\textsuperscript{1,*}\\[6pt]
\textsuperscript{1}Travel Statistics Bureau\\
\textsuperscript{*}Corresponding author: \href{mailto:contact@travelstatisticsbureau.com}{contact@travelstatisticsbureau.com}}
\date{September 2026\\[4pt]
{\small Preprint. An expanded version with interactive tables is at \url{https://travelstatisticsbureau.com/research/average-travel-time-to-work/}\\
Licensed CC BY 4.0. Travel Statistics Bureau, \url{https://travelstatisticsbureau.com/}}}

\begin{document}
\maketitle
\vspace{-1.2em}
\begin{abstract}
The mean travel time to work is one of the most cited statistics of the American Community Survey (ACS), yet its distribution, its dependence on mode and departure time and its relation to other measurement bases are rarely set out together. Using ACS 1-year tables, we report that US workers aged 16 and over who did not work from home averaged 27.2 minutes one way in 2024, against 25.0 minutes in 2006, a peak of 27.6 in 2019 and 25.6 in 2021. Dividing published aggregate minutes by workers, commuter rail, long-distance train and ferry riders averaged 71.0 minutes, 2.68 times the 26.5 minutes of solo drivers, and all public transportation 49.5 minutes. In 2024, 9.3\% of commuters, about 13.3 million people, travelled 60 minutes or more each way and 3.0\% 90 minutes or more; commuters leaving between 5:00 and 5:29 a.m. averaged 35.1 minutes. State means ranged from 33.2 minutes in New York to 17.5 in North Dakota, and big-metro means from 36.5 minutes in New York to 21.5 in Buffalo. Working from home rose from 5.7\% of workers in 2019 to 17.9\% in 2021 and stood at 13.3\% in 2024. The 2022 National Household Travel Survey recorded an average commute trip of 13.43 miles at 25.51 mph. On different definitions, the 2019 EU labour force survey module gave an EU average of 25 minutes and 30 minutes for the United Kingdom, and England's travel diary a commuting trip of 31.5 minutes in 2025 against 26.5 in 2002. The three bases are reported side by side and never pooled. All inputs are public.

\end{abstract}

\section{Introduction}

The mean travel time to work is among the most widely quoted statistics of the American Community Survey (ACS).
It is a single number for a very uneven quantity: most commuters reach work in well under half an hour, while a
minority travel for an hour or more each way, and the time depends strongly on the means of transport, the hour
of departure and the place of residence. The number also moved sharply during the pandemic years, when a large
share of workers stopped travelling to a workplace at all \citep{acs_s0801}.

This paper asks how long the journey to work takes in the United States, how that time is distributed across
commuters, modes, departure times and places, and how it has changed since 2006. We answer it with the ACS 1-year
tables on travel time, means of transportation and time of departure
\citep{acs_s0801,acs_b08303,acs_b08301,acs_b08136,acs_b08302,acs_b08133,census_mean}. Because the ACS records the
usual time of the trip but not its length, we add the distance and speed of commute trips from the National
Household Travel Survey (NHTS) \citep{nhts2022}. For an international view we report the 2019 European labour
force survey module on commuting \citep{eurostat} and the time per commuting trip from England's National Travel
Survey \citep{dft_nts0403}. The three sources use different definitions, and we keep them separate throughout.

The contribution is descriptive. We (i) report the 2024 national mean and its 2006--2024 path, by sex; (ii)
describe the distribution of commuters by one-way time band and the share with trips of 60 and 90 minutes or
more; (iii) derive mean trip times by mode and by departure time from published aggregate minutes; (iv) compare
all 50 states, the District of Columbia, Puerto Rico and the 51 metro areas with at least 500,000 commuters; (v)
summarise trip length, speed and daily travel from the travel diary; and (vi) set the US figures beside European
and English measures of the same activity.

\section{Data}

\subsection{American Community Survey}

The ACS asks each worker how many minutes it usually took to get from home to work in the reference week. The
Census Bureau publishes the mean of this quantity, for workers aged 16 and over who did not work from home, in
the subject table S0801 (Commuting Characteristics by Sex), which also gives the mean by sex, the distribution of
means of transportation for all workers and the shares of commuters in selected time bands \citep{acs_s0801}. We
use the national S0801 tables for 2010--2024 and the state tables for 2019 and 2024; the 2024 tables for metro and
micro areas give the same measures for 535 areas. National means for 2006--2009 come from the Census Bureau's
published chart of mean travel time to work \citep{census_mean}. The 1-year estimates for 2020 were released only
as experimental data and are not part of the official series, so 2020 is absent from every ACS series in this
paper.

The detailed tables add counts. B08303 gives the number of commuters in twelve one-way time bands, 2010--2024
\citep{acs_b08303}. B08301 gives the number of workers by means of transportation, including those who worked
from home \citep{acs_b08301}, and B08136 the aggregate travel time in minutes of commuters by the same means
\citep{acs_b08136}. B08302 gives the number of commuters by time of leaving home, in fourteen slots
\citep{acs_b08302}, and B08133 their aggregate travel time \citep{acs_b08133}. All ACS values are 1-year estimates
covering the twelve months of each calendar year.

\subsection{Travel diary}

The NHTS records the actual trips that members of sampled households made on an assigned travel day. The Federal
Highway Administration's summary of the 2022 survey reports the average length, duration and speed of commute
trips by mode for 2009, 2017 and 2022, the average commute speed by the population of the metropolitan statistical
area (MSA) for 2001--2022, the usual mode of travel to work for 2001--2022, and daily person trips and person-miles
of travel by purpose back to 1969 \citep{nhts2022}. Distances are network-calculated from 2009 (with the 2009
values adjusted) and were reported by respondents before that; transit time includes access and egress but not
waiting. The 2022 survey used a new method and a smaller sample.

\subsection{Europe and England}

European figures come from the one-off 2019 ad hoc module of the EU labour force survey (EU-LFS), table
lfso\_19plwk28 \citep{eurostat}. It asked employed people aged 15--74 how long their one-way trip from home to
their main job takes without detours, in whole minutes, and covers everyone with a trip of at least one minute.
We use the mean for the 27 EU countries, Iceland, Norway, Switzerland, the United Kingdom and T\"urkiye, and the
number of persons in each time band. England's figures come from the Department for Transport table NTS0403, which
reports, for each year 2002--2025, the number of commuting trips per person, the average time per commuting and per
education trip including waiting time, and the average trip length \citep{dft_nts0403}. The 2020--2022 values were
affected by changes to data collection during the pandemic, and in 2025 most diaries were completed digitally.

\section{Methods}

\subsection{Definitions}

For a group $g$ of commuters (a mode, or a departure slot) with published aggregate one-way travel time $A_g$ in
minutes and $N_g$ workers, the mean one-way travel time is
\begin{equation}
  \bar{t}_g = \frac{A_g}{N_g}.
\end{equation}
The ACS publishes $A_g$ and $N_g$ for modes (B08136, B08301) and departure slots (B08133, B08302) but not the
means, which are therefore our calculation. Applied to all commuters in 2024, the formula gives
$3{,}895{,}171{,}370 / 143{,}334{,}079 = 27.2$ minutes, identical to the published mean
\citep{acs_b08136,acs_b08303}.

The share of commuters in a set of time bands $B$ is $s_B = 100 \times \sum_{b \in B} n_b / \sum_b n_b$, with
$n_b$ the commuters in band $b$ of B08303. Mode shares divide the workers of each means of transportation by all
workers, including those who worked from home (B08301). For Europe, the share with trips of 60 minutes or more is
the number of persons in the 60-minute-and-over bands divided by all persons commuting one minute or more, from the
thousands of persons published by Eurostat. Differences between years are in minutes or in percentage points;
relative changes are $100 \times (x_{t_1}/x_{t_0} - 1)$.

A ``daily'' commute is taken to be twice the one-way mean, which assumes that the trip home takes as long as the
trip to work; the ACS measures only the outbound trip. Annual daily-travel distance is the daily person-miles
multiplied by 365. Miles are converted at 1.609344 km per mile.

\subsection{Trend lines}

Linear trends are ordinary least-squares fits of the annual value on the year over the stated range, with the
missing 2020 omitted for ACS series. We fit the US mean over 2006--2024 (18 points), the share of commuters under
10 minutes over 2010--2024 (14 points), and England's time per commuting and education trip over 2002--2025 (24
points), and report the fitted value for 2050 only as a descriptive extrapolation, not a forecast. No trend is
fitted to England's commuting trips per person, whose 2020 fall is a step change rather than a drift.

\subsection{Groupings and comparability}

States are the 50 states and the District of Columbia; Puerto Rico appears in the state table and is noted where
it would rank. Big metro areas are the 51 metropolitan areas with at least 500,000 commuters in 2024. The US
(usual one-way time of workers 16 and over), European (one-way time without detours of employed people 15--74)
and English (time per diary trip including waiting) measures are not comparable in level. They are reported in
separate figures and compared only as trends or approximate orders of magnitude.

\section{Results}

\subsection{The national mean and its path since 2006}

\begin{figure}[t]
\centering
\includegraphics[width=\textwidth]{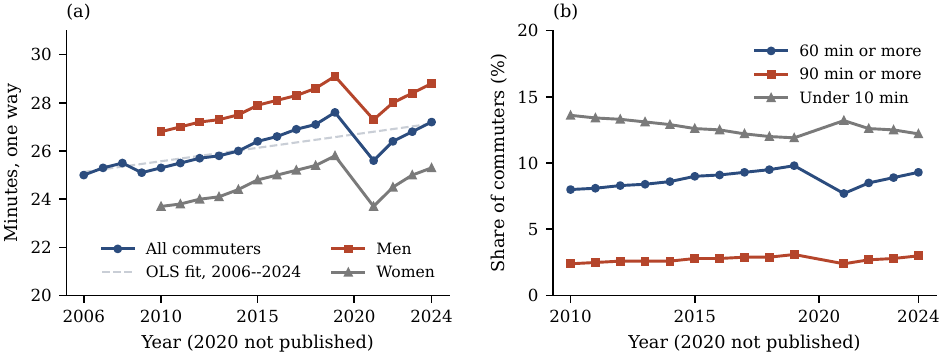}
\caption{(a) Mean one-way travel time to work, US, 2006--2024, all commuters and by sex (2010--2024), with the
least-squares line for all commuters. (b) Share of commuters with one-way trips of 60 minutes or more, 90 minutes
or more and under 10 minutes, 2010--2024. 2020 not published. Sources: \citet{census_mean,acs_s0801,acs_b08303}.}
\label{fig:trend}
\end{figure}

In 2024 the mean one-way travel time to work was 27.2 minutes, with a margin of error of 0.1 minute, for the
143.3 million workers who travelled to a workplace \citep{acs_s0801,acs_b08303}. Doubling it gives about 54.4
minutes of commuting per working day. The mean rose almost every year from 25.0 minutes in 2006 \citep{census_mean}
to a peak of 27.6 minutes in 2019, fell to 25.6 minutes in 2021 and recovered by 1.6 minutes, or 6.3\%, to 2024,
when it stood 0.4 minute below the peak and 2.2 minutes above 2006 (Figure~\ref{fig:trend}a). The least-squares
line over 2006--2024 rises 0.11 minute a year and reaches about 30.0 minutes in 2050.

Men commuted longer than women in every year. In 2024 men averaged 28.8 minutes and women 25.3, a gap of 3.5
minutes; in 2010 the two means were 26.8 and 23.7 minutes, and both fell in 2021 (to 27.3 and 23.7)
\citep{acs_s0801}.

\subsection{Distribution of trip times}

Most trips were short (Figure~\ref{fig:dist}a). In 2024, 61.3\% of commuters reached work in under 30 minutes;
the largest band, 15--19 minutes, held 15.0\% and the smallest, under 5 minutes, 2.8\% \citep{acs_b08303}. The
30--34 minute band held 13.7\%, against 7.0\% in the 25--29 band and 3.4\% in the 35--39 band; the spike is
consistent with respondents rounding their answer to half an hour rather than with a real cluster of trips. At the
upper end, 6.3\% of commuters travelled 60--89 minutes and 3.0\% 90 minutes or more, so that 9.3\%, about 13.3
million people, travelled an hour or more each way and about 4.3 million 90 minutes or more.

Long commutes followed the path of the mean (Figure~\ref{fig:trend}b). The share at 60 minutes or more rose from
8.0\% in 2010 to 9.8\% in 2019, fell to 7.7\% in 2021 and returned to 9.3\% in 2024; the share at 90 minutes or more
moved from 2.4\% to 3.1\%, back to 2.4\% and then to 3.0\% over the same years \citep{acs_b08303}. Short trips became
rarer: the share under 10 minutes fell from 13.6\% in 2010 to a low of 11.9\% in 2019, rose to 13.2\% in 2021 and was
12.2\% in 2024. Its least-squares line over 2010--2024 falls 0.08 point a year and reaches about 10.2\% in 2050.

\subsection{Means of transportation}

\begin{figure}[t]
\centering
\includegraphics[width=\textwidth]{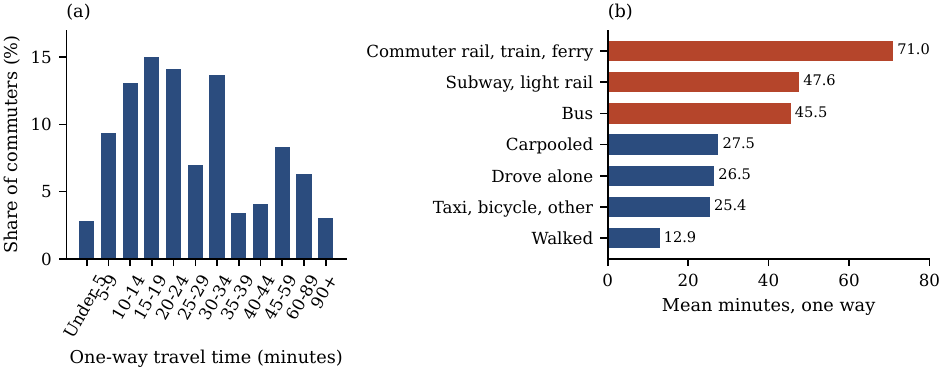}
\caption{(a) Commuters by usual one-way travel time band, US, 2024 (bands of unequal width: 45--59, 60--89 and 90+
minutes are wider). (b) Mean one-way travel time by means of transportation, 2024, computed as aggregate minutes
divided by workers; modes averaging over 40 minutes highlighted. Sources: \citet{acs_b08303,acs_b08136,acs_b08301}.}
\label{fig:dist}
\end{figure}

Of the 165.4 million workers in 2024, 69.2\% drove alone, 13.3\% worked from home, 9.2\% carpooled, 3.7\% used
public transportation and 4.6\% walked, cycled or used a taxi, ride-hailing, a motorcycle or other means
\citep{acs_b08301}. The mean one-way time differed by a factor of more than five across modes
(Figure~\ref{fig:dist}b). Commuter rail, long-distance train and ferry riders averaged 71.0 minutes, 2.68 times the
26.5 minutes of solo drivers; subway, light rail and streetcar riders averaged 47.6 minutes and bus riders 45.5.
Public transportation as a whole averaged 49.5 minutes, 1.87 times a solo drive. Carpoolers averaged 27.5 minutes,
the taxi, motorcycle, bicycle and other group 25.4, and walkers 12.9, a figure consistent with walking being chosen
mainly by people who live close to work \citep{acs_b08136,acs_b08301}.

The composition of modes changed more than the time of each trip (Table~\ref{tab:modes}). Working from home was
4.3\% of workers in 2010 and 5.7\% in 2019, jumped to 17.9\% in 2021 and settled at 13.3\% in 2024, 4.6 points below
2021 and 7.6 points above 2019 \citep{acs_s0801}. Driving alone fell from 75.9\% in 2019 to 69.2\% in 2024 and public
transportation from 5.0\% to 3.7\%. The number of commuters fell from 148.0 million in 2019 to 126.7 million in 2021
and was 143.3 million in 2024, still below the 2019 level.

\begin{table}[t]
\centering
\small
\caption{Means of transportation to work, US, 2010--2024: percent of all workers aged 16 and over, and commuters
(workers who did not work from home) in millions. 2020 not published. Sources: \citet{acs_s0801,acs_b08303}.}
\label{tab:modes}
\begin{tabular}{lrrrrrrr}
\toprule
Year & Drove alone & Carpooled & Public transp. & Walked & Bicycle & Worked from home & Commuters (M) \\
\midrule
2010 & 76.6 & 9.7 & 4.9 & 2.8 & 0.5 & 4.3 & 131.0 \\
2011 & 76.4 & 9.7 & 5.0 & 2.8 & 0.6 & 4.3 & 132.3 \\
2012 & 76.3 & 9.7 & 5.0 & 2.8 & 0.6 & 4.4 & 134.7 \\
2013 & 76.4 & 9.4 & 5.2 & 2.8 & 0.6 & 4.4 & 136.7 \\
2014 & 76.5 & 9.2 & 5.2 & 2.7 & 0.6 & 4.5 & 139.3 \\
2015 & 76.6 & 9.0 & 5.2 & 2.8 & 0.6 & 4.6 & 141.5 \\
2016 & 76.3 & 9.0 & 5.1 & 2.7 & 0.6 & 5.0 & 142.8 \\
2017 & 76.4 & 8.9 & 5.0 & 2.7 & 0.5 & 5.2 & 144.8 \\
2018 & 76.3 & 9.0 & 4.9 & 2.6 & 0.5 & 5.3 & 146.4 \\
2019 & 75.9 & 8.9 & 5.0 & 2.6 & 0.5 & 5.7 & 148.0 \\
2021 & 67.8 & 7.8 & 2.5 & 2.2 & 0.4 & 17.9 & 126.7 \\
2022 & 68.7 & 8.6 & 3.1 & 2.4 & 0.5 & 15.2 & 136.2 \\
2023 & 69.2 & 9.0 & 3.5 & 2.4 & 0.5 & 13.8 & 139.9 \\
2024 & 69.2 & 9.2 & 3.7 & 2.4 & 0.5 & 13.3 & 143.3 \\
\bottomrule
\end{tabular}
\end{table}

\subsection{Departure time}

\begin{figure}[t]
\centering
\includegraphics[width=\textwidth]{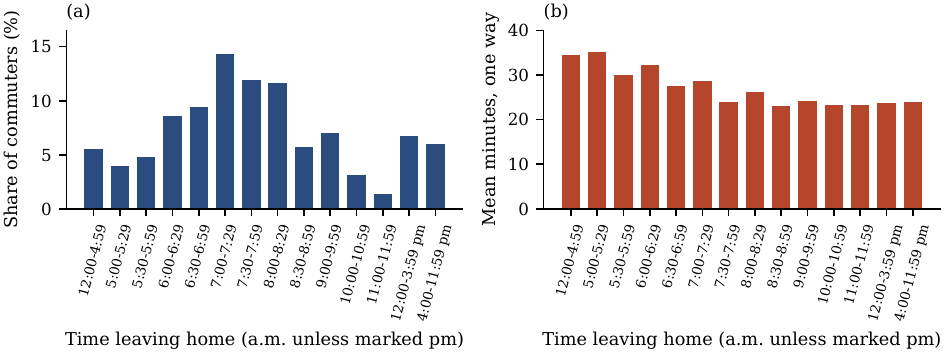}
\caption{(a) Share of commuters by time leaving home, US, 2024. (b) Mean one-way travel time by time leaving home,
computed as aggregate minutes divided by commuters in each slot. Sources: \citet{acs_b08302,acs_b08133}.}
\label{fig:dep}
\end{figure}

Commuters left home mostly between 6 and 9 a.m. (Figure~\ref{fig:dep}a). The busiest half hour was 7:00--7:29 a.m.,
when 14.3\% of commuters set off, and 37.8\% left between 7:00 and 8:29 a.m. \citep{acs_b08302}. Another 14.3\% left
before 6 a.m., including 5.5\% between midnight and 4:59 a.m.; the quietest slot was 11:00--11:59 a.m., with 1.4\%.

Mean trip time fell through the morning (Figure~\ref{fig:dep}b). Commuters leaving between 5:00 and 5:29 a.m.
averaged 35.1 minutes, the longest of any slot, and those leaving between midnight and 4:59 a.m. 34.4 minutes. Every
slot up to 6:29 a.m. averaged 30.0 minutes or more, while every slot from 7:30 a.m. onwards averaged under 27
minutes. The 7:00--7:29 a.m. peak averaged 28.7 minutes, 6.4 minutes less than the 5:00 group, and the 8:30--8:59
a.m. slot 23.1 minutes, one of the shortest \citep{acs_b08133,acs_b08302}.

\subsection{States}

State means in 2024 ranged from 33.2 minutes in New York to 17.5 minutes in North Dakota, a difference of 15.7
minutes; the New York mean was 1.90 times the North Dakota one (Table~\ref{tab:states}). Maryland (32.3) and New
Jersey (31.9) followed New York, and the District of Columbia averaged 31.0; South Dakota had the second-shortest
mean, 18.5 minutes \citep{acs_s0801}. The same three states led the share of commuters travelling 60 minutes or
more: 17.2\% in New York, 15.2\% in New Jersey and 14.5\% in Maryland. Short trips were most common in rural states:
32.6\% of commuters in Wyoming reached work in under 10 minutes, 30.0\% in North Dakota and 29.9\% in South Dakota.

Several of the longest commutes were shorter in 2024 than in 2019: New York by 0.8 minute (from 34.0), Maryland by
1.4 minutes (from 33.7) and New Jersey by 1.2 minutes (from 33.1). Minnesota, a short-commute state, fell by 1.7
minutes, from 24.4 to 22.7. Others lengthened, among them Mississippi, from 25.1 to 26.4 minutes, and Maine, from
24.6 to 25.6 \citep{acs_s0801}.

\begin{table}[t]
\centering
\small
\caption{Mean one-way travel time to work (minutes) and selected shares (percent), 2024, with the 2019 mean: the ten
longest and the ten shortest of the 50 states and the District of Columbia. Puerto Rico (28.6 minutes) would rank
ninth. Source: \citet{acs_s0801}.}
\label{tab:states}
\begin{tabular}{lrrrrrr}
\toprule
State & 2024 & 2019 & 60+ min & Under 10 min & Worked from home & Public transp. \\
\midrule
New York & 33.2 & 34.0 & 17.2 & 9.6 & 11.8 & 24.2 \\
Maryland & 32.3 & 33.7 & 14.5 & 8.0 & 15.6 & 5.0 \\
New Jersey & 31.9 & 33.1 & 15.2 & 9.5 & 13.5 & 9.7 \\
District of Columbia & 31.0 & 31.7 & 8.9 & 4.2 & 22.9 & 26.6 \\
Massachusetts & 30.6 & 31.0 & 13.3 & 10.1 & 14.8 & 7.8 \\
California & 29.7 & 30.7 & 12.1 & 9.1 & 14.1 & 3.3 \\
Georgia & 29.4 & 29.3 & 11.3 & 9.8 & 14.9 & 1.2 \\
Florida & 28.9 & 28.4 & 9.6 & 8.6 & 15.5 & 1.3 \\
Illinois & 28.3 & 29.6 & 10.4 & 11.8 & 13.6 & 7.0 \\
Virginia & 28.3 & 29.1 & 10.0 & 9.8 & 15.8 & 3.0 \\
\midrule
Iowa & 20.4 & 19.8 & 4.5 & 23.2 & 9.9 & 0.8 \\
Kansas & 20.0 & 20.0 & 4.0 & 22.1 & 10.3 & 0.3 \\
Nebraska & 19.9 & 19.5 & 4.0 & 22.4 & 9.7 & 0.5 \\
Alaska & 19.8 & 20.2 & 5.2 & 25.4 & 8.4 & 0.7 \\
Wyoming & 19.7 & 19.1 & 7.4 & 32.6 & 9.7 & 1.1 \\
Montana & 19.5 & 19.5 & 5.0 & 24.0 & 11.7 & 0.8 \\
South Dakota & 18.5 & 17.9 & 4.4 & 29.9 & 8.0 & 0.2 \\
North Dakota & 17.5 & 17.9 & 4.3 & 30.0 & 7.1 & 0.5 \\
\bottomrule
\end{tabular}
\end{table}

\subsection{Metro areas}

Among the 51 metro areas with at least 500,000 commuters, New York--Newark--Jersey City had the longest mean in 2024,
36.5 minutes, followed by Riverside--San Bernardino--Ontario (34.2), Washington--Arlington--Alexandria (33.8), San
Francisco--Oakland--Fremont (32.7) and Atlanta--Sandy Springs--Roswell (32.4) \citep{acs_s0801}. The two longest
reflect different mechanisms: 27.2\% of New York-area workers used public transportation, the highest share of the
group, against 0.9\% in Riverside--San Bernardino, so that long mean times arose from both slow transit and long
drives. The Chicago area averaged 30.9 minutes (margin of error 0.2), 3.7 minutes above the national mean, with
12.4\% of commuters travelling an hour or more, 9.0\% of workers on public transportation and 14.9\% working from home.

At the short end, Buffalo--Cheektowaga averaged 21.5 minutes, Grand Rapids--Wyoming--Kentwood 21.9 and
Milwaukee--Waukesha 23.2 (Table~\ref{tab:metros}). In Buffalo 3.3\% of commuters travelled an hour or more and 10.4\% of workers worked from home, and a Buffalo commuter saved 15.0 minutes each way against one in the
New York area. Across all 535 metro and micro areas, New York had the longest mean, with East Stroudsburg,
Pennsylvania, within its margin of error, and Great Falls, Montana, the shortest, 14.5 minutes, with a few small areas
such as Walla Walla, Washington, within its margin of error \citep{acs_s0801}.

\begin{table}[t]
\centering
\small
\caption{The ten longest and the ten shortest mean one-way commutes (minutes) among the 51 US metro areas with
500,000 or more commuters, 2024. Metro names shortened to the principal city. Source: \citet{acs_s0801}.}
\label{tab:metros}
\begin{tabular}{rlrlr}
\toprule
Rank & Longest & Mean & Shortest & Mean \\
\midrule
1 & New York, NY--NJ & 36.5 & Buffalo, NY & 21.5 \\
2 & Riverside, CA & 34.2 & Grand Rapids, MI & 21.9 \\
3 & Washington, DC--VA--MD--WV & 33.8 & Milwaukee, WI & 23.2 \\
4 & San Francisco, CA & 32.7 & Oklahoma City, OK & 23.5 \\
5 & Atlanta, GA & 32.4 & Salt Lake City, UT & 23.5 \\
6 & Boston, MA--NH & 31.9 & Kansas City, MO--KS & 23.7 \\
7 & Houston, TX & 31.1 & Cleveland, OH & 24.0 \\
8 & Chicago, IL--IN & 30.9 & Columbus, OH & 24.1 \\
9 & Los Angeles, CA & 30.7 & Minneapolis--St.~Paul, MN--WI & 24.1 \\
10 & Baltimore, MD & 30.3 & Hartford, CT & 24.3 \\
\bottomrule
\end{tabular}
\end{table}

\subsection{Distance and speed}

The travel diary measures the length and speed of actual commute trips (Table~\ref{tab:nhts}). In 2022 the average
commute trip covered 13.43 miles (21.6 km) and took 27.72 minutes, at an average speed of 25.51 mph (41.1 km/h)
\citep{nhts2022}. By private vehicle the trip averaged 13.56 miles (21.8 km) in 26.94 minutes. Transit trips were
shorter, 7.53 miles (12.1 km), but slower, 43.05 minutes at 10.68 mph (17.2 km/h). Walking commutes covered 1.28 miles
(2.1 km) in 25.05 minutes, and the average walking speed, on trips with both distance and time recorded, was 2.80 mph
(4.5 km/h). The 2017 diary recorded an average commute trip of 11.46 miles; because the 2022 survey used a new method,
the two years are separate snapshots rather than a trend.

\begin{table}[t]
\centering
\small
\caption{Average commute trip length, travel time and speed by mode from the National Household Travel Survey, 2009,
2017 and 2022. 2009 adjusted; distance network-calculated; transit time includes access and egress but not waiting;
2009 travel time not published. 2022 margins of error (all modes): length 1.30 mi, time 1.96 min, speed 0.92 mph.
Source: \citet{nhts2022}.}
\label{tab:nhts}
\begin{tabular}{llrrr}
\toprule
Mode & Year & Length (mi) & Time (min) & Speed (mph) \\
\midrule
All modes & 2009 & 11.03 & -- & 25.79 \\
 & 2017 & 11.46 & 26.58 & 23.42 \\
 & 2022 & 13.43 & 27.72 & 25.51 \\
Private vehicle & 2009 & 11.26 & -- & 27.01 \\
 & 2017 & 11.84 & 25.01 & 25.22 \\
 & 2022 & 13.56 & 26.94 & 26.59 \\
Public transit & 2009 & 10.18 & -- & 11.42 \\
 & 2017 & 12.09 & 58.11 & 11.63 \\
 & 2022 & 7.53 & 43.05 & 10.68 \\
Walk & 2009 & 0.98 & -- & 4.76 \\
 & 2017 & 1.19 & 15.26 & 3.15 \\
 & 2022 & 1.28 & 25.05 & 2.80 \\
\bottomrule
\end{tabular}
\end{table}

Commutes were slowest in the largest metro areas. In 2022 the average commute speed in MSAs of 3 million people or
more was 23.2 mph (37.3 km/h), against 31.4 mph (50.5 km/h) in rural areas outside any MSA; in 2001 the corresponding
figures were 25.4 and 31.9 mph \citep{nhts2022}. As published, the 2022 values for every MSA size class except rural
areas are identical to the 2017 values, so no change between those two years is read from them. The car remained the
usual way to work for 90.8\% of workers in 2001 and 90.5\% in 2022; transit reached 6.9\% in 2017 and fell to 4.1\% in
2022, while walking stayed between 2.8\% and 2.9\%.

Commuting is a minority of all travel. Counting every trip, the average person travelled 28.55 miles (46 km) a day in
2022, down from 38.67 in 1995, or about 10,400 miles (16,800 km) a year \citep{nhts2022}. Trips to and from work
accounted for 5.61 miles a day, 19.6\% of the total, against 11.43 miles for social and recreational trips; work
miles per person peaked at 8.69 a day in 1995, while shopping and errands fell from 13.51 to 5.79 miles between 1995
and 2022. Daily trips per person fell from 4.30 in 1995 to 2.28 in 2022. Since the 2022 method changed and distances
before 2009 were self-reported, these values are best read as snapshots.

\subsection{Europe}

\begin{figure}[t]
\centering
\includegraphics[width=\textwidth]{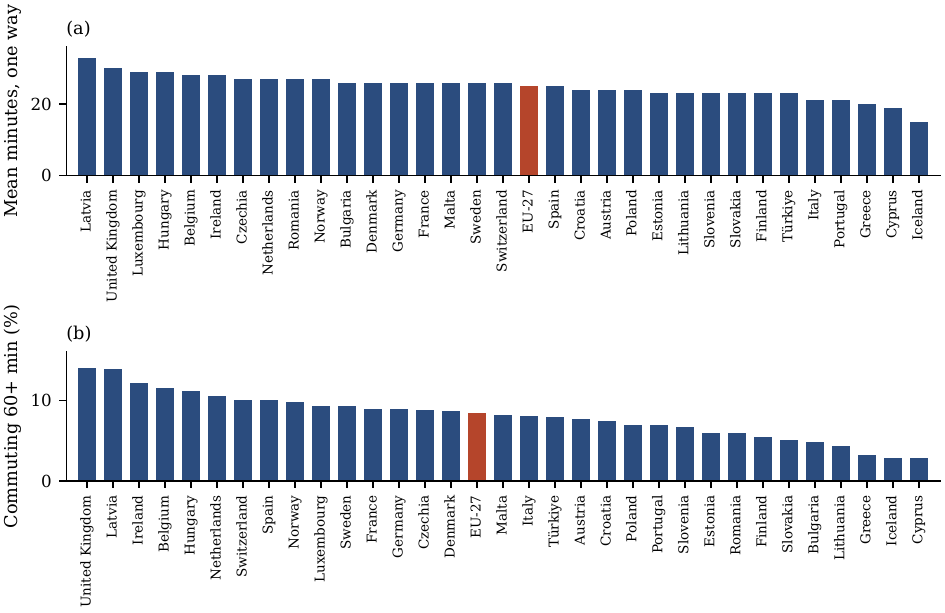}
\caption{(a) Mean one-way commuting time from home to work without detours, employed people aged 15--74 with a commute
of at least 1 minute, 2019, whole minutes; the 27 EU countries, Iceland, Norway, Switzerland, the United Kingdom and
T\"urkiye; EU-27 average highlighted. (b) Share of those commuters with a one-way trip of 60 minutes or more,
computed from thousands of persons; each panel ranked separately. Source: \citet{eurostat}.}
\label{fig:eu}
\end{figure}

In the 2019 EU-LFS module the mean one-way commute across the 27 EU countries was 25 minutes (Figure~\ref{fig:eu}a)
\citep{eurostat}. Latvia had the longest mean, 33 minutes, followed by the United Kingdom (30), Hungary and Luxembourg
(29 each) and Belgium and Ireland (28 each); Iceland had the shortest, 15 minutes, 18 minutes less than Latvia, with
Cyprus at 19 and Greece at 20. EU commuters living in cities averaged 27 minutes, against 24 in towns, suburbs and
rural areas. The share of commuters with trips of 60 minutes or more was highest in the United Kingdom (14.1\%),
Latvia (13.9\%) and Ireland (12.2\%), against 8.5\% for the EU as a whole and 2.8\% in Cyprus (Figure~\ref{fig:eu}b).
The US mean for the same year, 27.6 minutes on the ACS definition \citep{acs_s0801}, was about 3 minutes above the EU
average and about 2 minutes below the United Kingdom, a comparison of orders of magnitude only, since the populations
and questions differ.

\subsection{England}

\begin{figure}[t]
\centering
\includegraphics[width=\textwidth]{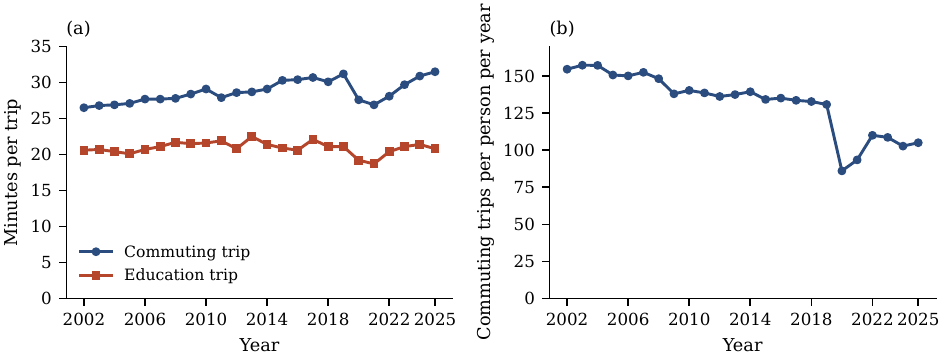}
\caption{England, 2002--2025. (a) Average minutes per commuting and per education trip, including waiting time, all
modes. (b) Commuting trips per person per year, excluding short walks. 2020--2022 affected by pandemic data
collection; 2025 diaries mostly digital. Source: \citet{dft_nts0403}.}
\label{fig:eng}
\end{figure}

England's travel diary measures time per commuting trip, including waiting, over all commuting trips recorded on the
diary days. On that basis a commuting trip took 31.5 minutes in 2025, 5.0 minutes more than the 26.5 minutes of 2002
(Figure~\ref{fig:eng}a) \citep{dft_nts0403}. The series rose to 31.2 minutes in 2019, fell to 26.9 in 2021 and has risen
in each year since. The least-squares line over 2002--2025 rises 0.15 minute a year and reaches about 34.1 minutes in
2050. The length of the commuting trip changed less, from 8.52 miles in 2002 to 9.19 miles (14.8 km) in 2025, a rise of
7.9\%, so the average commuting trip in England became slower.

The number of commuting trips fell. Commuting trips per person were 154.7 a year in 2002 and 130.9 in 2019, dropped to
86.1 in 2020 and were 105.1 in 2025, about 50 fewer than in 2002 (Figure~\ref{fig:eng}b). Education trips barely
changed: 20.8 minutes in 2025 against 20.6 in 2002, over 2.90 miles (4.7 km) in 2025, and a least-squares line that
reaches about 20.4 minutes in 2050. Measured on the same per-trip basis, the education trip was about 10.7 minutes
shorter than the commuting trip in 2025 \citep{dft_nts0403}.

\section{Discussion}

Three regularities stand out. First, the US commute lengthened slowly and steadily before 2020 and then returned close
to its previous level even though far fewer people travelled. In 2024 the mean was 27.2 minutes, 0.4 minute below the
2019 peak, while the share of workers at home was 13.3\%, 7.6 points above 2019, and 4.7 million fewer people
commuted than in 2019. Home working removed a large number of trips but left the typical remaining trip about as long
as before. The travel diary's 2022 commute trip of 13.43 miles, against 11.46 in 2017, points the same way, although
the change of survey method prevents a firm comparison \citep{nhts2022}.

Second, the longest commutes arise from two distinct sources. Public transportation trips averaged 49.5 minutes, and
New York, the state with the longest mean at 33.2 minutes, had 24.2\% of its workers on public transportation. Yet
car-based Riverside--San Bernardino, where 0.9\% used transit, had the second-longest big-metro mean, 34.2 minutes.
Long average times are therefore consistent both with a high transit share and with long distances driven, and state
or metro rankings alone do not separate the two.

Third, the time of departure matters more than the rush hour itself. The longest mean trips belonged to the earliest
departures, 35.1 minutes for the 5:00--5:29 a.m. group against 28.7 for the 7:00--7:29 a.m. peak. This is consistent
with long-distance commuters leaving earlier, not with congestion at the peak making trips longer; the tables do not
link departure time to distance, so the mechanism cannot be established here.

Internationally, the level of the US mean lies within the European range, and England's per-trip measure grew faster
over its period, 5.0 minutes between 2002 and 2025, than the US usual one-way time, 2.2 minutes between 2006 and 2024.
The two are different measures and are compared only as trends.

\section{Limitations}

The ACS measure is a usual one-way time reported by the worker, not a timed trip, and it covers only the outbound
journey; the daily figures assume a symmetrical return. Answers are rounded, which inflates the 30--34 minute band. The
mode and departure-time means are our division of published aggregates and inherit their sampling error; margins of
error are not reported for them. The 2020 ACS estimates are absent, and the 2021 values reflect an unusual year. State
and metro comparisons do not control for distance, income or occupation. The travel diary measures actual trips on a
single day with a different sample, its 2022 method is new, and its MSA speed table repeats the 2017 values in five of
six columns as published. The European figures come from a one-off 2019 module with a different age range and
question, and England's figures are per-trip times including waiting, affected by pandemic data collection in
2020--2022 and by digital diaries in 2025; neither is comparable in level with the ACS. Trend lines are descriptive
least-squares fits and are not forecasts.

\section{Data and code availability}

All inputs are public: the ACS 1-year tables S0801, B08303, B08301, B08136, B08302 and B08133
\citep{acs_s0801,acs_b08303,acs_b08301,acs_b08136,acs_b08302,acs_b08133} and the Census Bureau's mean travel time chart
\citep{census_mean}; the FHWA summary of the 2022 National Household Travel Survey \citep{nhts2022}; Eurostat table
lfso\_19plwk28 \citep{eurostat}; and the Department for Transport table NTS0403 \citep{dft_nts0403}. Every figure in
this paper, with its source and computation, is published with interactive tables at
\url{https://travelstatisticsbureau.com/research/average-travel-time-to-work/} under a CC BY 4.0 licence.

\section*{Acknowledgements}
The authors are volunteers with Travel Statistics Bureau, a not-for-profit project. No external funding was received.
Correspondence: \href{mailto:contact@travelstatisticsbureau.com}{contact@travelstatisticsbureau.com}. Website: \url{https://travelstatisticsbureau.com/}.

\bibliographystyle{plainnat}
\bibliography{refs}

\end{document}